\documentclass[a4paper,12pt]{article}
\usepackage{amsfonts,amsthm,amsmath,amssymb,graphicx}
\usepackage{braket}
\usepackage{color}
\begin{document}

\newtheorem{definition}{Definition}[section]
\newcommand{\be}{\begin{equation}}
\newcommand{\ee}{\end{equation}}
\newcommand{\bea}{\begin{eqnarray}}
\newcommand{\eea}{\end{eqnarray}}
\newcommand{\LE}{\left[}
\newcommand{\R}{\right]}
\newcommand{\nn}{\nonumber}
\newcommand{\Tr}{\text{Tr}}
\newcommand{\N}{\mathcal{N}}
\newcommand{\G}{\Gamma}
\newcommand{\vf}{\varphi}
\newcommand{\LL}{\mathcal{L}}
\newcommand{\Op}{\mathcal{O}}
\newcommand{\HH}{\mathcal{H}}
\newcommand{\arctanh}{\text{arctanh}}
\newcommand{\up}{\uparrow}
\newcommand{\down}{\downarrow}
\newcommand{\rd}{\partial}
\newcommand{\de}{\partial}
\newcommand{\ba}{\begin{eqnarray}}
\newcommand{\ea}{\end{eqnarray}}
\newcommand{\db}{\bar{\partial}}
\newcommand{\we}{\wedge}
\newcommand{\ca}{\mathcal}
\newcommand{\lr}{\leftrightarrow}
\newcommand{\f}{\frac}
\newcommand{\s}{\sqrt}
\newcommand{\vp}{\varphi}
\newcommand{\hvp}{\hat{\varphi}}
\newcommand{\tvp}{\tilde{\varphi}}
\newcommand{\tp}{\tilde{\phi}}
\newcommand{\ti}{\tilde}
\newcommand{\ap}{\alpha}
\newcommand{\pr}{\propto}
\newcommand{\mb}{\mathbf}
\newcommand{\ddd}{\cdot\cdot\cdot}
\newcommand{\no}{\nonumber \\}
\newcommand{\la}{\langle}
\newcommand{\lb}{\rangle}
\newcommand{\ep}{\epsilon}
 \def\we{\wedge}
 \def\lr{\leftrightarrow}
 \def\f {\frac}
 \def\ti{\tilde}
 \def\ap{\alpha}
 \def\pr{\propto}
 \def\mb{\mathbf}
 \def\ddd{\cdot\cdot\cdot}
 \def\no{\nonumber \\}
 \def\la{\langle}
 \def\lb{\rangle}
 \def\ep{\epsilon}
\def\m{{\mu}}
 \def\w{{\omega}}
 \def\n{{\nu}}
 \def\ep{{\epsilon}}
 \def\d{{\delta}}
 \def\rh{\rho}
 \def\t{{\theta}}
 \def\a{{\alpha}}
 \def\T{{\Theta}}
 \def\frac#1#2{{#1\over #2}}
 \def\l{{\lambda}}
 \def\G{{\Gamma}}
 \def\D{{\Delta}}
 \def\g{{\gamma}}
 \def\s{\sqrt}
 \def\ch{{\chi}}
 \def\b{{\beta}}
 \def\CA{{\cal A}}
 \def\CC{{\cal C}}
 \def\CI{{\cal I}}
 \def\CO{{\cal O}}
 \def\o{{\rm ord}}
 \def\Ph{{\Phi }}
 \def\L{{\Lambda}}
 \def\CN{{\cal N}}
 \def\p{\partial}
 \def\pslash{\p \llap{/}}
 \def\Dslash{D \llap{/}}
 \def\Mp{m_{{\rm P}}}
 \def\apm{{\alpha'}}
 \def\r{\rightarrow}
 \def\Re{{\rm Re}}
 \def\MG{{\bf MG:}}
\def\be{\begin{equation}}
\def\ee{\end{equation}}
\def\ba{\begin{eqnarray}}
\def\ea{\end{eqnarray}}
\def\bal{\begin{align}}
\def\eal{\end{align}}

 \def\de{\partial}
 \def\db{\bar{\partial}}
 \def\we{\wedge}
 \def\lr{\leftrightarrow}
 \def\f {\frac}
 \def\ti{\tilde}
 \def\ap{\alpha}
  \def\al{\alpha'}
 \def\pr{\propto}
 \def\mb{\mathbf}
 \def\ddd{\cdot\cdot\cdot}
 \def\no{\nonumber \\}
\def\nn{\nonumber \\}
 \def\la{\langle}
 \def\lb{\rangle}
 \def\ep{\epsilon}
 \def\ddbp{\mbox{D}p-\overline{\mbox{D}p}}
 \def\ddbt{\mbox{D}2-\overline{\mbox{D}2}}
 \def\ov{\overline}
 \def\cl{\centerline}
 \def\vp{\varphi}
\def\hB{\hat \Box}

\begin{titlepage}

\thispagestyle{empty}

\begin{flushright}
KUNS-2583\\
YITP-15-95\\
IPMU15-0197\\
\end{flushright}

\vspace{.4cm}
\begin{center}
\noindent{\large \textbf{Double-Trace Deformations and Entanglement Entropy in AdS}}\\
\vspace{2cm}

Taiki Miyagawa$^{a}$, Noburo Shiba$^{b}$ and Tadashi Takayanagi$^{b,c}$
\vspace{1cm}

{\it
$^{a}$Department of Physics, Kyoto University, Kyoto 606-8502, Japan\\
$^{b}$Yukawa Institute for Theoretical Physics (YITP),\\
Kyoto University, Kyoto 606-8502, Japan\\
$^{c}$Kavli Institute for the Physics and Mathematics of the Universe (Kavli IPMU),\\
University of Tokyo, Kashiwa, Chiba 277-8582, Japan\\
}

\vskip 2em
\end{center}

\vspace{.5cm}
\begin{abstract}
We compute the bulk entanglement entropy of a massive scalar field in a Poincare AdS with the Dirichlet and Neumann boundary condition when we trace out a half space. Moreover, by taking into account the quantum back reaction to the minimal surface area, we calculate how much the entanglement entropy changes under a double-trace deformation of a holographic CFT. In the AdS3/CFT2 setup, our result agrees with the known result in 2d CFTs. In higher dimensions, our results offer holographic predictions.
\end{abstract}

\end{titlepage}

\newpage

\section{Introduction}

The AdS/CFT correspondence \cite{Ma,GKP,Wa} offers us a powerful way to compute useful physical quantities
in strongly coupled quantum field theories in the large $N$ limit. One such example is the holographic calculation of entanglement entropy \cite{RT}. Until now, there have been
extensive works on the holographic entanglement entropy in the classical gravity limit (see e.g. references in \cite{Review}). However, explicit computations of quantum corrections dual to the $1/N$ expansions, pioneered by the works \cite{BDHM,FLM}, have been limited to several specific examples. Among them, the quantum corrections play a crucial role in the entanglement entropy and mutual information \cite{BDHM,FLM} for multi intervals \cite{He,Ha,Fa} in two dimensional CFTs and for its higher dimensional counterparts \cite{Shiba,Cardy,AF} (see also further related works \cite{Bin,DD,Pe,CSZ,BeMa,ShibaE,NaNi,HMPZ} and for other aspects of
quantum corrections to holographic entanglement entropy refer to \cite{SHS,SwRa,EW,Le,Kelly:2015mna} ).

The main purpose of this paper is to report a modest progress in a new example where the quantum corrections give the leading contributions. We will especially consider a setup of AdS/CFT correspondence which describes a double-trace deformation of a holographic CFT \cite{Wit,GM,GK}. We simply choose the half space to be the subsystem $A$ to define the entanglement entropy $S_A$.\footnote{The double trace deformation is also  used to construct a holographic model Kondo effect in \cite{ErHoOBWu} and its entanglement  entropy was analyzed in \cite{ErFlHoNeWu} in a classical gravity limit.}
In its gravity dual, we consider a massive scalar in an AdS space and compute its bulk entanglement entropy $S^{bulk}_{A}$, which is defined by the entanglement entropy for the bulk subsystem $\Sigma_A$, a plane ended on $A$ at the AdS boundary. This bulk entanglement entropy is closely related to the one-loop quantum correction $S^{1-loop}_A$ to
the holographic entanglement entropy, owing to the remarkable formula found by Faulkner, Lewkowycz and Maldacena \cite{FLM}. We are interested in how the entanglement entropy changes under the renormalization group flow when the double-trace deformation leads to a relevant perturbation. In the gravity description this change corresponds to the shift of boundary conditions from the Neumann to Dirichlet. We will show that even though $S^{bulk}_{A}$ includes bulk UV divergences, which are interpreted as the area law \cite{BKLS,Sr}
in an AdS space, the difference of entanglement entropy between the two boundary conditions are free from UV divergences.

One way to calculate the quantum corrections of the holographic entanglement entropy is to employ the replica method and to directly perform the path-integral in quantum gravity or string theory. Though such an analysis is in general quite difficult even for the leading corrections (i.e. one-loop corrections), this computation can be done analytically in AdS$_3/$CFT$_2$ \cite{BDHM}. Another way to calculate the leading quantum corrections to the holographic entanglement entropy is to employ the formula obtained in \cite{FLM} based on the generalized entropy calculation \cite{LM}. In this paper we employ the latter approach as we want to deal with higher dimensional AdS spaces. This prescription tells us that the one-loop quantum correction $S^{1-loop}_A$ is given by the sum of
the bulk entanglement $S^{bulk}_{A}$ and the change of minimal surface area $\f{{\large \delta}\mbox{Area}}{4G_N}$ due to the back reactions from the quantum energy stress tensor as well as some other contributions.
We will explicitly evaluate these contributions and obtain $S^{1-loop}_A$ in our
holographic setup.

This paper is organized as follows. In section two, we give a brief review of holographic description of double-trace deformations and entanglement entropy. In section three we compute the entanglement entropy in the bulk AdS space with a massive scalar field. In section four, we calculate the quantum corrections to the holographic entanglement entropy dual to a double-trace deformation of a holographic CFT. In section five we summarize our conclusions.

\section{Double-Trace Deformation and Entanglement Entropy}

Consider a scalar operator $O(x)$ in a $d+1$ dimensional CFT (CFT$_{d+1}$) dual to a scalar field $\phi(x,z)$ in a $d+2$ dimensional AdS space (AdS$_{d+2}$). In a holographic CFT described by a gauge theory, $O(x)$ should be gauge invariant and is typically given by a single trace operator $O(x)=\mbox{Tr}[\ddd]$, where $\ddd$ denotes a product of colored operators.

We focus on the Euclidean Poincare AdS whose metric takes the familiar form:
\be
ds^2=\f{R^2}{z^2}(dz^2+dx_0^2+\sum_{i=1}^{d} dx_i^2). \label{pads}
\ee

The mass $m$ of the scalar field is related to the conformal dimension of the scalar operator $O(x)$ via
the standard formula \cite{GKP,Wa,KW}:
\be
\Delta_{\pm}=\f{d+1}{2}\pm \nu,\ \ \ \left(\nu\equiv \s{\f{(d+1)^2}{4}+m^2R^2}\right). \label{dome}
\ee
By solving the equation of motion for the scalar field,  $\phi$ behaves like
\be
\phi(z,x)\simeq \alpha (x)z^{\Delta_-}+\beta(x)z^{\Delta_+},
\ee
in the near boundary limit $z\to 0$.
We define the Dirichlet and Neumann boundary condition by $\ap(x)=0$ and $\beta(x)=0$, respectively.
When the scalar field $\phi(z,x)$ obeys the Dirichlet and Neumann boundary condition at the AdS boundary $z=0$, the dual scalar operator $O(x)$ has the conformal dimension $\Delta_+$ and
$\Delta_{-}$, respectively \cite{KW}.  It is very useful to note that the shift
of boundary condition corresponds to the sign flip $\nu\to -\nu$ for our later analysis.

\subsection{Double-Trace Deformation}

Both of the two boundary conditions are sensible only in the range
$0\leq \nu\leq 1$. Indeed, starting from the Neumann boundary condition (we call its dual CFT$^{(N)}$), we find that the double-trace deformation of the CFT action $I_{CFT}$
by the dimension $2\Delta_{-}$ operator $O(x)^2$:
\be
\delta I_{CFT}=\lambda \int dx^{d+1}O(x)^2,  \label{double}
\ee
becomes relevant or marginal, with the unitarity bound $\Delta\geq \f{d-1}{2}$ satisfied.
Under the double-trace deformation (\ref{double}), it flows into another fixed point.
This new CFT (we call its dual CFT$^{(D)}$) corresponds to the Dirichlet boundary condition of the dual scalar field \cite{Wit}. The dimension of $O(x)^2$ becomes $2\Delta_+(>d+1)$ and thus it becomes irrelevant.

\subsection{Entanglement Entropy}

The main aim of this paper is to calculate the entanglement entropy for CFT$^{(N)}$ and CFT$^{(D)}$ defined on the flat spacetime $R^{d+1}$ by using their gravity duals. We define its cartesian coordinate by $(x_0,x_1,x_2,\ddd,x_d)$, where
$x_0$ denotes the Euclidean time.
To define the entanglement entropy, we first divide the $d$ dimensional space $R^d$ into two parts $A$ and $B$, which leads to the Hilbert space factorization $H=H_A\otimes H_B$.
Then we define the reduced density matrix $\rho_A=\mbox{Tr}[|\Psi\lb\la\Psi|]$ by tracing out $H_B$ , where $|\Psi\lb$ is the ground state. The entanglement entropy $S_A$ is defined as
\be
S_A=-\mbox{Tr}[\rho_A\log \rho_A],
\ee
where $\rho_A$ is the reduced density matrix. We can express this quantity as a derivative
\be
S_A=\lim_{n\to 1}\left[-\f{\de}{\de n}\log \mbox{Tr}[\rho_A^n]\right]. \label{rep}
\ee

We choose the subsystem $A$ to be a half of the flat space $R^d$ defined by $x_1\geq 0$. If we define the polar coordinate such that $x_1+ix_0=\rho e^{i\theta}$, then the trace $\mbox{Tr}[\rho_A^n]$ is given by the ratio $\f{Z_n}{(Z_1)^n}$,
where $Z_n$ is the partition function of a given CFT for the $n$-sheeted space defined by $0\leq \theta \leq 2\pi n$. We first assume that
$n$ is a positive integer and later analytically continue $n$ to take the derivative, which is
called the replica method.

In this paper we focus on the behavior of entanglement entropy when we perform a double-trace deformation (\ref{double}) of a holographic CFT. In the classical gravity or tree level limit (i.e. $O(G_N^{-1})$), the holographic entanglement entropy \cite{RT} dual to $S_A$ defined in the above setting is given by
\be
S^{tree}_A=\f{\mbox{Area}(\gamma_A)}{4G_N},
\ee
where $\gamma_A$ is the minimal surface which ends on $\de A$ in the AdS boundary and is simply given by $d$ dimensional plane $x_1=x_0=0$ in the Poincare AdS spacetime (\ref{pads}). The non-compact volume in the
$x^2\sim x^d$ direction is expressed as $V_{d-1}$ in this paper.

However, since double-trace deformations only affect the gravity dual partition function at one loop order or higher, the classical contributions are the same both for CFT$^{(N)}$ and CFT$^{(D)}$ as given by
(for $d\geq 2$)
\be
S^{tree}_A=\f{R^d}{4G_N}\int d^{d-1}x\int^\infty_{\ep}\f{dz}{z^{d}}
=\f{V_{d-1}}{(d-1)\ep^{d-1}}\cdot \f{R^d}{4G_N}, \label{heed}
\ee
where $\ep$ is the near boundary cut off in AdS dual to the UV cut off in CFT. For $d=1$, we need a IR cut off $z_{IR}$, interpreted as the correlation length and we get
\be
S^{tree}_A=\f{R}{4G_N}\int^{z_{IR}}_{\ep}\f{dz}{z}=\f{c}{6}\log\f{z_{IR}}{\ep}, \label{heeone}
\ee
which agrees with the well-known CFT result \cite{HLW,CC}. Here we employed the familiar relation between the central charge of two dimensional CFT and the AdS radius \cite{BrHe}.

Therefore, in order to distinguish CFT$_N$ and CFT$_D$,  we need to study the first order quantum corrections to holographic entanglement entropy. As derived in \cite{FLM} by Faulkner, Lewkowycz and Maldacena, the holographic entanglement entropy at one loop level (i.e. $O(G_N^0)$) consists of contributions from the bulk entanglement entropy $S^{bulk}_A$, the shift of minimal surface area $\f{\delta \mbox{Area}}{4G_N}$ due to the quantum corrections, the Wald-like entropy $S_{Wald}$ and the counter terms $S_{c.t.}$:
\be
S^{1-loop}_A=S^{bulk}_A+\f{\delta \mbox{Area}}{4G_N}+S_{Wald}+S_{c.t.}. \label{loopee}
\ee
In our holographic model, we assume that $\phi$ is the minimally coupled scalar field in AdS and we do not turn on interaction terms in the scalar field action which lead to the extra Wald-like entropy such as the curvature coupling $\int R\phi^2$. Therefore we can ignore $S_{Wald}$. The counter terms $S_{c.t.}$ simply cancels bulk UV divergences in the first two terms. We write the bulk UV cut off
as $\delta$, while another infinitesimally small parameter $\ep$ denotes the UV cut off in CFTs.

\section{Entanglement Entropy in AdS}

 One essential part of holographic entanglement entropy at one loop level (\ref{loopee}) is equal to the entanglement entropy in the bulk: $S^{bulk}_A$. Thus in this section we would like to compute the entanglement entropy of a scalar field theory on AdS$_{d+2}$ by applying the replica calculation (\ref{rep}). This quantity is also interesting because it offers an analytical example of entanglement entropy in a curved space.

The subsystem for this bulk entanglement entropy is chosen to be the $d+1$ dimensional half plane $\Sigma_A$ defined by $x_1>0$ and $x_0=0$ in the Poincare AdS spacetime (\ref{pads}). Note that the boundary of this subsystem is given by the previous minimal surface $\gamma_A$ defined by $x_1=x_0=0$.

\subsection{Entanglement Entropy in AdS from Orbifolds}

Instead of taking $n$ to be an positive integer, we set $n=1/N$ with $N$ chosen to be a positive integer \cite{NiTa,HNTW}. This trick simplifies our presentation of computations. However, straightforward calculations in appendix A show that this actually leads to the same result as that from the standard replica method. Using this trick, we have $\mbox{Tr}[\rho_A^n]=\f{Z^{AdS}_{1/N}}{(Z^{AdS}_1)^{1/N}}$, where
$Z^{AdS}_{1/N}$ is the partition function on the orbifold AdS$_{d+2}/{\mbox{Z}_N}$. The Z$_N$ orbifold action is defined to be $x_1+ix_0\to e^{\f{2\pi}{N}i}(x_1+ix_0)$.
Thus the entanglement entropy in AdS can be computed as
\be
S^{bulk}_A=\lim_{N\to 1}\f{\de}{\de N}\left[\log Z^{AdS}_{1/N}-\f{1}{N}\log Z^{AdS}_1\right]. \label{repp}
\ee

We would like to compute the contribution to the holographic entanglement entropy $S_A$ from a scalar field on Euclidean $AdS_{d+2}$. The action is given by
\begin{equation}
\begin{split}
I_{G}=\int d^{d+2} x \sqrt{g} \dfrac{1}{2} ( \nabla^{a} \phi \nabla_{a} \phi +m^2 \phi^2 ) ,
\end{split}  \label{action}
\end{equation}
where $a$ runs the integers $0,1,\ddd,d+1$ and the metric is given by (\ref{pads}).

We consider the following eigenvalue problem for the Laplacian,
\begin{equation}
\begin{split}
\left(-R^2\cdot \Box-\dfrac{(d+1)^2}{4} \right)\phi = \alpha^2 \phi
\end{split}  \label{eigenvalue eq}
\end{equation}
where $\Box= \frac{1}{\sqrt{g}} \partial_{a} (\sqrt{g} g^{ab} \partial_{b}) $. The parameter
$\alpha^2$ is defined by $m^2R^2+\dfrac{(d+1)^2}{4}+\ap^2=0$ and is from now on
regarded as an eigenvalue.

This eigenvalue problem can be solved with the ansatz \cite{Cal}
\begin{equation}
\begin{split}
\phi =e^{i\vec{p} \cdot \vec{x}} (pz)^{\frac{d+1}{2}} f(pz) ,
\end{split}  \label{ansatz}
\end{equation}
where $\vec{x}=(x_0, x_1, \cdots , x_d)$ and $p=|\vec{p}|$.
We substitute (\ref{ansatz}) into (\ref{eigenvalue eq}) and obtain
\begin{equation}
\begin{split}
0&=\left(\Box + \frac{1}{R^2} (\alpha^2+\dfrac{(d+1)^2}{4} ) \right)\phi \\
&=e^{i\vec{p} \cdot \vec{x}} \frac{1}{R^2} \left( u^2 f''(u) +u f'(u) -(u^2-\alpha^2 ) f \right),
\end{split}  \label{eigenvalue eq2}
\end{equation}
where $u\equiv pz$.
The solutions of  (\ref{eigenvalue eq2}) are $f(u)=K_{i\alpha}(u), I_{i\alpha}(u)$.
By requiring the solutions to be well-behaved at $z\rightarrow \infty$,
we choose $f(u)=K_{i\alpha}(u)$.

By imposing the boundary condition for the direction $(x_0, x_1)$,
we obtain the eigenvalue function as
\begin{equation}
\begin{split}
\braket{x|\alpha, \vec{p}} \equiv \phi_{\alpha, \vec{p}} (\vec{x} , z) = F_{\vec{p}} (\vec{x}) (pz)^{\frac{d+1}{2}} K_{i\alpha}(pz) ,
\end{split}  \label{eigenfunction 1}
\end{equation}
where
\begin{equation}
\begin{split}
 F_{\vec{p}} (\vec{x}) =\frac{1}{(2\pi)^{(d+1)/2}N} e^{i\vec{p_{\perp}} \cdot \vec{x_{\perp}}}
 \sum_{j=0}^{N-1} e^{i(g^j \vec{p_{\parallel}}) \cdot \vec{x_{\parallel}}},
\end{split}  \label{eigenfunction 2}
\end{equation}
here $\vec{x_{\perp}}=(x_2, x_3, \cdots , x_d)$ and  $\vec{x_{\parallel}}=(x_0, x_1)$,
and $g$ is a rotational operator, i.e.
\begin{equation}
\begin{split}
g \vec{x_{\parallel}}=(\cos \theta x_0-\sin \theta  x_1, \sin \theta x_0 +\cos \theta x_1)
\end{split}  \label{rotation}
\end{equation}
where $\theta=2\pi/N$.
$ F_{\vec{p}} (\vec{x}) $ satisfy the boundary condition
$ F_{\vec{p}} (g \vec{x_{\parallel}}, \vec{x_{\perp}} ) =  F_{\vec{p}} (\vec{x_{\parallel}}, \vec{x_{\perp}} ) $
and are normalized as
\begin{equation}
\begin{split}
\int d^{d+1} x  F^*_{\vec{p}} (\vec{x}) F_{\vec{p'}} (\vec{x})=\frac{1}{N}
 \sum_{j=0}^{N-1} \delta^2 (\vec{p_{\parallel}}- g^j \vec{p'_{\parallel}})
 \cdot \delta^{d-1} (\vec{p_{\perp}}-  \vec{p'_{\perp}}) ,
\end{split}  \label{normalization 1 }
\end{equation}
and
\begin{equation}
\begin{split}
\int d^{d+1} p  F^*_{\vec{p}} (\vec{x}) F_{\vec{p}} (\vec{x'})
=\frac{1}{N}
 \sum_{j=0}^{N-1} \delta^2 (\vec{x_{\parallel}}- g^j \vec{x'_{\parallel}})
 \cdot \delta^{d-1} (\vec{x_{\perp}}-  \vec{x'_{\perp}}) .
\end{split}  \label{normalization 2}
\end{equation}

Thus $\braket{x|\alpha, \vec{p}}$ is normalized as
\begin{equation}
\begin{split}
&\int d^{d+2} x \sqrt{ g(x) } \braket{\alpha, \vec{p}|x } \braket{x|\alpha', \vec{p'}} \\
&= \frac{1}{N}
 \sum_{j=0}^{N-1} \delta^2 (\vec{x_{\parallel}}- g^j \vec{x'_{\parallel}})
 \cdot \delta^{d-1} (\vec{x_{\perp}}-  \vec{x'_{\perp}})
 \cdot R^{d+2} p^{d+1} (\mu (\alpha))^{-1} \delta(\alpha-\alpha')  ,
\end{split}  \label{normalization 3}
\end{equation}
where
\begin{equation}
\begin{split}
\mu (\alpha)= \frac{2 \alpha}{\pi^2} \sinh \pi \alpha .
\end{split}  \label{measure}
\end{equation}
In (\ref{normalization 3}) we have used (see Appendix \ref{ap:or})
\begin{equation}
\begin{split}
\int_0^{\infty} dz \frac{1}{z} K_{i\alpha}(z)  K_{i\alpha'}(z) =(\mu (\alpha))^{-1} \delta(\alpha-\alpha') .
\end{split}  \label{orthogonal}
\end{equation}

Thus, by introducing the bulk UV cut off $\delta$  we obtain
\begin{equation}
\begin{split}
\ln Z^{AdS}_{1/N}&=\frac{1}{2} \int_{\delta^2}^{\infty} \frac{ds}{s} \mbox{Tr}e^{-s(-\Box+m^2)} \\
&= \frac{1}{2} \int_{\delta^2}^{\infty} \frac{ds}{s}
\int d^{d+2}x \sqrt{ g(x) }
 \int_0^{\infty} d \alpha \mu (\alpha) \int d^{d+1}p \frac{1}{R^{d+2} p^{d+1}}  \\
& ~~~ \times  e^{-\frac{s}{R^2} (\alpha^2+\nu^2) }
  \braket{x|\alpha, \vec{p}}  \braket{\alpha, \vec{p}|x }    \\
& =   \frac{1}{2} \int_{\delta^2}^{\infty} \frac{ds}{s} \int_\epsilon^{\infty} dz \frac{1}{z}
 \int_0^{\infty} d \alpha \mu (\alpha) \int d^{d+1}p
 \frac{1}{N}
 \sum_{n=0}^{N-1} \delta^2 (\vec{p_{\parallel}}- g^j \vec{p_{\parallel}})
 \cdot \delta^{d-1} (\vec{p_{\perp}}-  \vec{p_{\perp}})            \\
& ~~~ \times  e^{-\frac{s}{R^2} (\alpha^2+\nu^2) }
   K_{i\alpha}(pz)  K_{i\alpha}(pz)   \\
\end{split}  \label{partition}
\end{equation}
where
\begin{equation}
\begin{split}
\nu^2=m^2 R^2+ \frac{(d+1)^2}{4} .
\end{split}
\end{equation}
The delta function is evaluated as
\begin{equation}
\begin{split}
\sum_{j=0}^{N-1} \delta^2 (\vec{p_{\parallel}}- g^j \vec{p_{\parallel}})
 \cdot \delta^{d-1} (\vec{p_{\perp}}-  \vec{p_{\perp}})
 =\frac{1}{N} \left( \frac{V_2}{(2\pi)^2} + \delta (\vec{p_{\parallel}}) \cdot \frac{N^2-1}{12} \right)
 \cdot \frac{V_{d-1}}{(2\pi)^{d-1}},
\end{split}  \label{delta}
\end{equation}
where we employed the formula
\be
\sum_{j=1}^{N-1}\f{1}{\sin^2\f{\pi j}{N}}=\f{N^2-1}{3}.
\ee

We substitute (\ref{delta}) into (\ref{partition}) and obtain
\begin{equation}
\begin{split}
&\ln Z^{AdS}_{1/N}- \frac{1}{N} \ln Z^{AdS}_1    \\
& =   \frac{1}{2}\cdot \frac{N^2-1}{12N} \frac{V_{d-1}}{(2\pi)^{d-1}}
\int_{\delta^2}^{\infty} \frac{ds}{s} \int_\epsilon^{\infty} dz \frac{1}{z}
 \int_0^{\infty} d \alpha \mu (\alpha) \int d^{d-1} \vec{p_{\perp}}           \\
& ~~~ \times  e^{-\frac{s}{R^2} (\alpha^2+\nu^2) }
   K_{i\alpha}(p_{\perp} z)  K_{i\alpha}(p_{\perp}z)   \\
& =   \frac{1}{2}\cdot \frac{N^2-1}{12N} \frac{V_{d-1}}{(2\pi)^{d-1}} V(S_{d-2})
\int_{\delta^2}^{\infty} \frac{ds}{s} \int_\epsilon^{\infty} dz \frac{1}{z}
 \int_0^{\infty} d \alpha \mu (\alpha) \int_0^{\infty} d p_{\perp} ~ p_{\perp}^{d-2}            \\
& ~~~ \times  e^{-\frac{s}{R^2} (\alpha^2+\nu^2) }
   K_{i\alpha}(p_{\perp} z)  K_{i\alpha}(p_{\perp}z)   \\
& =   \frac{1}{2}\cdot \frac{N^2-1}{12N} \frac{V_{d-1}}{(2\pi)^{d-1}} V(S_{d-2})
\frac{\pi^{1/2} \Gamma \left(\frac{d-1}{2}\right) }{4 \Gamma (d/2) }
\int_{\delta^2}^{\infty} \frac{ds}{s} \int_\epsilon^{\infty} dz \frac{1}{z}
 \int_0^{\infty} d \alpha \mu (\alpha)             \\
& ~~~ \times  e^{-\frac{s}{R^2} (\alpha^2+\nu^2) }
  \frac{1}{z^{d-1}} \Gamma(\rho_d-i\alpha)  \Gamma(\rho_d+i\alpha)  ,
\end{split}  \label{partition 2}
\end{equation}
where $\rho_d\equiv \frac{d-1}{2}$. In the final expression in (\ref{partition 2}) we assumed
$d\geq 2$.
We use the following formula,
\begin{equation}
\begin{split}
\Gamma(\rho_d-i\alpha)  \Gamma(\rho_d+i\alpha) \mu(\alpha) =\frac{2}{\pi} q_{d}(\alpha)
\end{split}  \label{formula}
\end{equation}
where
\begin{equation}
\begin{split}
q_{d} (\alpha) = \begin{cases}
\prod_{j=0}^{\rho_d -1} (\alpha^2+j^2)  & \text{when $d$ is odd and $d\geq3$ }   \\
\alpha \tanh \pi \alpha \prod_{j=1/2}^{\rho_d -1} (\alpha^2+j^2) & \text{when $d$ is even and $d\geq2$} .
\end{cases}
\end{split}
\label{q_d-1}
\end{equation}
We expand $q_{d} (\alpha) $ as
\begin{equation}
\begin{split}
q_{d} (\alpha) = \begin{cases}
\sum_{n=1}^{\rho_d } c_{2n} \alpha^{2n}  & \text{when $d$ is odd and $d\geq3$ }   \\
 \tanh \pi \alpha \sum_{n=0}^{d/2 -1} c_{2n+1} \alpha^{2n+1}  & \text{when $d$ is even  and $d\geq2$ } .
\end{cases}
\end{split} \label{q_d-2}
\end{equation}
We substitute (\ref{formula}) into (\ref{partition 2}) and obtain
\begin{equation}
\begin{split}
&\ln Z_{1/N}- \frac{1}{N} \ln Z_1    \\
& =   \frac{1}{2}\cdot \frac{N^2-1}{12N} \frac{V_{d-1}}{(2\pi)^{d-1}} V(S_{d-2})
\frac{\pi^{1/2} \Gamma \left(\frac{d-1}{2}\right) }{4 \Gamma (d/2) }
\int_{\delta^2}^{\infty} \frac{ds}{s} \int_\epsilon^{\infty} dz \frac{1}{z^d}
 e^{-\frac{s}{R^2} \nu^2 }  \\
& ~~~ \times \int_0^{\infty} d \alpha  e^{-\frac{s}{R^2} \alpha^2 }
  \frac{2}{\pi} q_{d}(\alpha)  .
\end{split}  \label{partition 3}
\end{equation}
Note that the integral of $s$ in the above only includes the divergences from the bulk UV limit $\delta \to 0$.

\subsubsection{Detailed Evaluation: The case $d$ is odd and $d\geq3$}
In this case we can perform the $\alpha$ integral and obtain
\begin{equation}
\begin{split}
&\ln Z^{AdS}_{1/N}- \frac{1}{N} \ln Z^{AdS}_1    \\
& =   \frac{1}{2}\cdot \frac{N^2-1}{12N} \frac{V_{d-1}}{(2\pi)^{d-1}} V(S_{d-2})
\frac{\pi^{1/2} \Gamma \left(\frac{d-1}{2}\right) }{4 \Gamma (d/2) }
\int_{\delta^2}^{\infty} \frac{ds}{s} \int_\epsilon^{\infty} dz \frac{1}{z^d}
 e^{-\frac{s}{R^2} \nu^2 }  \\
& ~~~ \times  \frac{1}{\pi} \sum_{n=1}^{\rho_d} c_{2n} \left( \frac{s}{R^2} \right)^{-\frac{1}{2}-n} \Gamma (1/2+n)  \\
& =   \frac{1}{2}\cdot \frac{N^2-1}{12N} \frac{V_{d-1}}{(2\pi)^{d-1}} V(S_{d-2})
\frac{\pi^{1/2} \Gamma \left(\frac{d-1}{2}\right) }{4 \Gamma (d/2) }
\frac{1}{d-1}  \frac{1}{\epsilon^{d-1}}
  \\
& ~~~ \times  \frac{1}{\pi} \sum_{n=1}^{\rho_d} c_{2n} \Gamma (1/2+n)
\nu^{1+2n} [\Gamma (-1/2-n) -  \gamma(-1/2-n, (\nu \delta / R)^2 )]\\
& =   \frac{1}{2}\cdot \frac{N^2-1}{12N} \frac{V_{d-1}}{(2\pi)^{d-1}} V(S_{d-2})
\frac{\pi^{1/2} \Gamma \left(\frac{d-1}{2}\right) }{4 \Gamma (d/2) }
\frac{1}{d-1}  \frac{1}{\epsilon^{d-1}}
  \\
& ~~~ \times \left[ -2 \int_0^{\nu} d\tilde{\nu} q_d(i \tilde{\nu}) +  \frac{1}{\pi} \sum_{n=1}^{\rho_d} c_{2n} \Gamma (1/2+n) (\delta / R)^{-1-2n} \sum_{m=0}^{\infty} \dfrac{(-1)^m }{m! (1/2+n-m)}
\left( \dfrac{\nu \delta}{R} \right)^{2m} \right]  ,   \\
\end{split}  \label{partition odd}
\end{equation}
where $\gamma(s,x) \equiv \int_0^x dt t^{s-1} e^{-t}$ is the lower incomplete gamma function and
we expanded  $\gamma(s,x)$ and used (\ref{q_d-2}).
Thus we obtain the bulk entanglement entropy in AdS as
\begin{equation}
\begin{split}
&S^{bulk}_A=
-\frac{\partial}{\partial(1/N)} ( \ln Z^{AdS}_{1/N}- \frac{1}{N} \ln Z^{AdS}_1)|_{N=1}    \\
& =  \frac{1}{12} \frac{V_{d-1}}{(2\pi)^{d-1}} V(S_{d-2})
\frac{\pi^{1/2} \Gamma \left(\frac{d-1}{2}\right) }{4 \Gamma (d/2) }
\frac{1}{d-1}  \frac{1}{\epsilon^{d-1}}
  \\
& ~~~ \times \left[ -2 \int_0^{\nu} d\tilde{\nu} q_d(i \tilde{\nu}) +  \frac{1}{\pi} \sum_{n=1}^{\rho_d} c_{2n} \Gamma (1/2+n) (\delta / R)^{-1-2n} \sum_{m=0}^{\infty} \dfrac{(-1)^m }{m! (1/2+n-m)}
\left( \dfrac{\nu \delta}{R} \right)^{2m} \right] \\
    & \sim   \frac{1}{12} \frac{V_{d-1}}{(2\pi)^{d-1}} V(S_{d-2})
\frac{ \Gamma \left(\frac{d-1}{2}\right) }{2 \sqrt{\pi} }
\cdot \frac{1}{d} \left( \frac{R}{\delta} \right)^{d}
\cdot \frac{1}{d-1}  \frac{1}{\epsilon^{d-1}}+O(\delta^{-d+2}) .   \\
\end{split}  \label{ee oddd}
\end{equation}

This leading divergence can be understood as the bulk area law as it is proportional to
\be
\f{\mbox{Vol}(\gamma_A)}{\delta^{d}}=\f{R^d}{\delta^d} \int^\infty_\ep \f{dz}{z^d}
=\f{R^dV_{d-1}}{(d-1)\ep^{d-1}\delta^d},
\ee
where remember that $\gamma_A$ is the $d$ dimensional minimal surface defined by $x_0=x_1=0$.
At the same time, note that the fact that all terms in $S^{bulk}_A$ is proportional to $\f{V_{d-1}}{\ep^{d-1}}$ is consistent with the area law of the dual CFT \cite{BKLS,Sr}. Also notice that the bulk divergent terms in the limit $\delta\to 0$ are not affected by the boundary condition we take as is clear from the fact that it is symmetric under $\nu\to -\nu$.

On the other hand, the finite contributions depend on the boundary conditions.
The difference between Neumann and Dirichlet boundary condition
$S^{bulk}_A(-\nu) - S^{bulk}_A(\nu)$ is given by
\begin{equation}
\begin{split}
&S^{bulk}_A(-\nu) - S^{bulk}_A(\nu)
 =  \frac{1}{12} \frac{V_{d-1}}{(2\pi)^{d-1}} V(S_{d-2})
\frac{\pi^{1/2} \Gamma \left(\frac{d-1}{2}\right) }{ \Gamma (d/2) }
\frac{1}{d-1}  \frac{1}{\epsilon^{d-1}}
 \int_0^{\nu} d\tilde{\nu} q_d(i \tilde{\nu}).
  \\
\end{split}  \label{ee odd}
\end{equation}

\subsubsection{Detailed Evaluation: The case $d=1$}


For $d=1$, we substitute (\ref{delta}) into (\ref{partition}) and obtain
\begin{equation}
\begin{split}
&\ln Z^{AdS}_{1/N}- \frac{1}{N} \ln Z^{AdS}_1    \\
& =   \frac{1}{2}\cdot \frac{N^2-1}{12N}
\int_{\delta^2}^{\infty} \frac{ds}{s} \int_\epsilon^{\infty} dz \frac{1}{z}
 \int_0^{\infty} d \alpha \mu (\alpha) \int d^{2} \vec{p_{\parallel}}
 \delta^2(\vec{p_{\parallel}})
 e^{-\frac{s}{R^2} (\alpha^2+\nu^2) }
   K_{i\alpha}(p_{\parallel} z)  K_{i\alpha}(p_{\parallel} z) .  \\
\end{split}  \label{partition 2 d1}
\end{equation}
When $p_{\parallel}z\to 0$, we have
\begin{equation}
\begin{split}
&(K_{i\alpha} (p_{\parallel} z))^2 \sim \dfrac{\pi}{2\alpha \sinh (\pi \alpha)} -\dfrac{\pi^2}{4 (\sinh (\pi \alpha))^2}
\left[ \dfrac{(p_{\parallel} z/2)^{-2i \alpha} }{ (\Gamma (-i\alpha +1))^2}
- \dfrac{(p_{\parallel} z/2)^{2i \alpha} }{ (\Gamma (i\alpha +1))^2}  \right] .
\end{split}  \label{bessel K}
\end{equation}
The terms in the square bracket oscillate rapidly when  $pz \to 0$,
so we can neglect the contribution from these terms in (\ref{partition 2 d1})
and obtain
\begin{equation}
\begin{split}
&\ln Z^{AdS}_{1/N}- \frac{1}{N} \ln Z^{AdS}_1    \\
& =   \frac{1}{2}\cdot \frac{N^2-1}{12N}
\int_{\delta^2}^{\infty} \frac{ds}{s} \int_\epsilon^{\infty} dz \frac{1}{z}
 \int_0^{\infty} d \alpha \mu (\alpha)    e^{-\frac{s}{R^2} (\alpha^2+\nu^2) }
 \dfrac{\pi}{2\alpha \sinh (\pi \alpha)}     \\
 & =   \frac{1}{2}\cdot \frac{N^2-1}{12N}
\dfrac{1}{2 \sqrt{\pi}} \int_{\delta^2/R^2}^{\infty}  \frac{ds}{s^{3/2}} e^{-s \nu^2} \int_\epsilon^{\infty} dz \frac{1}{z}    \\
 & =   \frac{1}{2}\cdot \frac{N^2-1}{12N}
\dfrac{1}{2 \sqrt{\pi}}
\left[\nu \Gamma (-1/2)
-\dfrac{R}{ \delta} \sum_{m=0}^\infty \dfrac{(-1)^m (\delta^2 \nu^2 /R^2)^m}{m! (-1/2+m)} \right]
  \int_\epsilon^{\infty} dz \frac{1}{z}  .  \\
\end{split}  \label{partition 3 d1}
\end{equation}
Thus we obtain the bulk entanglement entropy in $AdS_3$ as
\begin{equation}
\begin{split}
&S^{bulk}_A=
-\frac{\partial}{\partial(1/N)} ( \ln Z^{AdS}_{1/N}- \frac{1}{N} \ln Z^{AdS}_1)|_{N=1}    \\
& =  \dfrac{1}{12}
\left[-\nu
-\dfrac{R}{2\sqrt{\pi} \delta} \sum_{m=0}^\infty \dfrac{(-1)^m (\delta^2 \nu^2 /R^2)^m}{m! (-1/2+m)} \right]
  \int_\epsilon^{\infty} dz \frac{1}{z}  .  \\
\end{split}  \label{ee oddd d1}
\end{equation}
The difference between Neumann and Dirichlet boundary condition
$S^{bulk}_A(-\nu) - S^{bulk}_A(\nu)$ is given by
\begin{equation}
\begin{split}
&S^{bulk}_A(-\nu) - S^{bulk}_A(\nu)
 =  \dfrac{\nu}{6}  \int_\epsilon^{\infty} dz \frac{1}{z} .
\end{split}  \label{eedd1pp}
\end{equation}

\subsubsection{Detailed Evaluation: The case $d=2$}
As an example of even $d$ case, let us study the entanglement entropy for $d=2$ below.
We will present general results for even $d$ in the next subsection.
For $d=2$, the equation (\ref{partition 2}) becomes
\begin{equation}
\begin{split}
&\ln Z^{AdS}_{1/N}- \frac{1}{N} \ln Z^{AdS}_1    \\
& =   \frac{1}{2}\cdot \frac{N^2-1}{12N} \frac{V_{1}}{2\pi}
 \int_\epsilon^{\infty}  \frac{dz}{z^2}
\int_{\delta^2}^{\infty} \frac{ds}{s}
 \int_0^{\infty} d \alpha  \alpha \tanh (\pi \alpha)          e^{-\frac{s}{R^2} (\alpha^2+\nu^2) }  .
\end{split}  \label{partition 2 d2}
\end{equation}
We perform the $s$ and the $\alpha$ integrals
in the similar way as
the calculation of the partition function in $AdS_2$ in \cite{Cal},
\begin{equation}
\begin{split}
&\int_{\delta^2}^{\infty} \frac{ds}{s}
 \int_0^{\infty} d \alpha  \alpha \tanh (\pi \alpha)          e^{-\frac{s}{R^2} (\alpha^2+\nu^2) }  \\
&=\int_{\delta^2/R^2}^{\infty} \frac{ds}{s}
 \int_0^{\infty} d \alpha \left( \alpha -\dfrac{2\alpha}{e^{2\pi \alpha}+1} \right)
     e^{-s (\alpha^2+\nu^2) }  \\
&=\int_{\delta^2/R^2}^{\infty} \frac{ds}{s}
\left( \dfrac{1}{2s} -2H_0 \right) e^{-s \nu^2 }
-2H_0 \ln \nu^2
+2 \int_0^{\infty} d \alpha \dfrac{\alpha}{e^{2\pi \alpha}+1} \ln (\alpha^2+\nu^2),   \\
\end{split}  \label{integral d2}
\end{equation}
where
\begin{equation}
\begin{split}
H_0\equiv \int_0^{\infty} d \alpha \dfrac{\alpha}{e^{2\pi \alpha}+1}=\dfrac{1}{48},
\end{split}  \label{H0 d2}
\end{equation}
and we used
\begin{equation}
\int_{\delta^2/R^2}^{\infty} \frac{ds}{s}
    ( e^{-s (\alpha^2+\nu^2) } -  e^{-s \nu^2 } )= \ln \nu^2 - \ln (\alpha^2+\nu^2)
+O(\delta^2/R^2)
\end{equation}
and we neglected terms which vanish when $\delta^2/R^2 \to 0$ in (\ref{integral d2}).
The last term in the right hand side of (\ref{integral d2}) is computed as
\begin{equation}
\begin{split}
 \int_0^{\infty} d \alpha \dfrac{\alpha}{e^{2\pi \alpha}+1} \ln (\alpha^2+\nu^2)
&= \int_0^{\infty} d \alpha \dfrac{\alpha}{e^{2\pi \alpha}+1} \ln \alpha^2
+ \int_0^{\nu^2} d\tilde{\nu}^2
\int_0^{\infty} d \alpha
 \dfrac{\alpha}{e^{2\pi \alpha}+1} \dfrac{1}{\alpha^2+\tilde{\nu}^2}  \\
&=\dfrac{1}{24} (\ln 2+12 \zeta'(-1))
+\dfrac{1}{2} \int_0^{\nu^2} d\tilde{\nu}^2
 ( \psi (\tilde{\nu} +1/2) -\ln \tilde{\nu} )  \\
&=\dfrac{1}{24} (\ln 2+12 \zeta'(-1))
+\dfrac{\nu^2}{4} -\dfrac{\nu^2}{2} \ln \nu
+ \int_0^{\nu} d\tilde{\nu}
 \tilde{\nu} \psi (\tilde{\nu} +1/2) , \\
\end{split}  \label{integral2 d2}
\end{equation}
where $\zeta(x)$ is the Riemann zeta function and $\psi (x)=\frac{d}{dx} \ln \Gamma(x)$ is the digamma function.
We substitute (\ref{integral2 d2}) into (\ref{integral d2}) and,
after some calculation,
we obtain
\begin{equation}
\begin{split}
&\int_{\delta^2}^{\infty} \frac{ds}{s}
 \int_0^{\infty} d \alpha  \alpha \tanh (\pi \alpha)          e^{-\frac{s}{R^2} (\alpha^2+\nu^2) }  \\
&=\dfrac{R^2}{2\delta^2} +\left( \nu^2 +4H_0 \right)
\left( \ln (\delta/R) +\gamma/2 \right)
+\dfrac{1}{12} (\ln 2+12 \zeta'(-1))
+2 \int_0^{\nu} d\tilde{\nu}
 \tilde{\nu} \psi (\tilde{\nu} +1/2),  \\
\end{split}  \label{integral3 d2}
\end{equation}
where $\gamma$ is the Euler's constant 
and we neglected terms which vanish when $\delta^2/R^2 \to 0$.
From (\ref{partition 2 d2}) and (\ref{integral3 d2}), we obtain
the bulk entanglement entropy in $AdS_4$ as
\begin{equation}
\begin{split}
&S^{bulk}_A=
-\frac{\partial}{\partial(1/N)} ( \ln Z^{AdS}_{1/N}- \frac{1}{N} \ln Z^{AdS}_1)|_{N=1}    \\
& = \frac{1}{12} \frac{V_{1}}{2\pi}
\left(\dfrac{R^2}{2\delta^2} +\left( \nu^2 +\dfrac{1}{12} \right)
\left( \ln (\delta/R) +\gamma/2 \right)
+\dfrac{1}{12} (\ln 2+12 \zeta'(-1))
+2 \int_0^{\nu} d\tilde{\nu}
 \tilde{\nu} \psi (\tilde{\nu} +1/2) \right)   \\
& ~~~~ \times \int_\epsilon^{\infty} \frac{dz}{z^2} , \\
\end{split}  \label{ee d2}
\end{equation}
where we used $H_0=1/48$.
The difference between Neumann and Dirichlet boundary condition
$S^{bulk}_A(-\nu) - S^{bulk}_A(\nu)$ is given by
\begin{equation}
\begin{split}
S^{bulk}_A(-\nu) - S^{bulk}_A(\nu)
& =  \frac{V_{1}}{12\pi}
 \int_0^{\nu} d\tilde{\nu}
 \tilde{\nu} (\psi (-\tilde{\nu} +1/2)- \psi (\tilde{\nu} +1/2) )
  \int_\epsilon^{\infty}  \frac{dz}{z^2}  \\
&= - \frac{V_{1}}{12}
 \int_0^{\nu} d\tilde{\nu}
 \tilde{\nu} \tan (\pi \tilde{\nu} )
  \int_\epsilon^{\infty}  \frac{dz}{z^2}  .
\end{split}  \label{eed2pp}
\end{equation}

\subsection{Entanglement Entropy in AdS from Green Functions}

We can equivalently compute the difference of entanglement entropy in AdS for Neumann and Dirichlet boundary condition from Green functions in AdS space.

The scalar field Green function in Poincare AdS$_{d+2}$ (see e.g. \cite{GM}) is given by
\ba
\la x,z|(\Box-m^2)^{-1}|y,w\lb &=&-R^{-d}(zw)^{\f{d+1}{2}}\int \f{dk^{d+1}}{(2\pi)^{d+1}}e^{ik(x-y)}I_\nu(kz)K_{\nu}(kw)\no
&=&-R^{-d}(zw)^{\f{d+1}{2}}\int \f{dk^{d+1}}{(2\pi)^{d+1}}e^{ik(x-y)}\int^\infty_{0}\f{ds}{2s}e^{-sk^2}
e^{-\f{z^2+w^2}{4s}}I_\nu\left(\f{zw}{2s}\right).\ \ \ \ \ \label{grfd}
\ea

By using this we can express the derivative of relevant partition function with respect to the mass square as follows:
\ba
&& \f{\de}{\de m^2}\left[\log Z_{AdS/Z_N}-(1/N)\log Z_{AdS}\right] \no
&& =-\f{1}{2}\cdot\f{\de}{\de m^2} \mbox{Tr}\left[\f{\sum^{N-1}_{j=1}g^j}{N}\log (\Box-m^2)\right] \no
&& =\f{1}{2}\mbox{Tr}\left[\f{\sum^{N-1}_{j=1}g^j}{N}
\f{1}{\Box-m^2}\right]\no
&&=-\f{R^2 V_{d-1}}{2(2\pi)^{d-1}}\cdot \f{N^2-1}{12N}\int^\infty_\ep\f{dz}{z}\int dk^{d-1}\int^\infty_{\delta^2}\f{ds}{2s} e^{-sk^2}e^{-\f{z^2}{2s}}I_\nu\left(\f{z^2}{2s}\right).\no
\ea

The derivative of entanglement entropy with respect to $m^2$ is given by
\be
\f{\de S^{bulk}_A}{\de \nu^2}=R^{-2}\f{\de S^{bulk}_A}{\de m^2}=-\f{V_{d-1}}{12}\int^\infty_\ep \f{dz}{z}\int^\infty_{\delta^2}
\f{ds}{2s}\cdot (4\pi s)^{-\f{d-1}{2}}e^{-\f{z^2}{2s}}I_\nu\left(\f{z^2}{2s}\right),
\label{delm}
\ee
Note that $S^{bulk}_A(\nu)$ is the bulk entanglement entropy in AdS for a scalar field
(mass $mR=\s{\nu^2-\left(\f{d+1}{2}\right)^2}$) with the Dirichlet boundary condition. The
 entanglement entropy for the same scalar field with the Neumann boundary condition is given by $S^{bulk}_A(-\nu)$. Note that the Green function expression (\ref{grfd}) at $k=0$ includes divergences for $d=1,2$ if we replace $\nu$ with $-\nu$ to calculate $S^{bulk}_A(-\nu)$ (remember we assume $\nu\in [0,1]$). Thus our argument in this subsection can only be applied to the case $d\geq 3$.

To see the universal finite part we can consider a difference $S_A(\nu)-S_A(0)$, which is evaluated as
\ba
S^{bulk}_A(\nu)-S^{bulk}_A(0)=-\f{V_{d-1}}{12} \int^\infty_\ep \f{dz}{z}\int^\infty_{\delta^2}
\f{ds}{2s}\cdot (4\pi s)^{-\f{d-1}{2}}e^{-\f{z^2}{2s}}\int^{\ti{\nu}^2}_0 d\ti{\nu}^2 I_\nu\left(\f{z^2}{2s}\right).
\ea

The difference between Neumann and Dirichlet boundary condition
$S^{bulk}_A(-\nu)-S^{bulk}_A(\nu)$ is given by as follows (we assume $d\geq 3$)
\ba
&& S^{bulk}_A(-\nu)-S^{bulk}_A(\nu) \no
&&=-\f{V_{d-1}}{12} \int^\infty_\ep \f{dz}{z}\int^\infty_{0}
\f{ds}{2s}\cdot (4\pi s)^{-\f{d-1}{2}}e^{-\f{z^2}{2s}}\int^{\nu^2}_0 d\ti{\nu}^2 \left(I_{-\ti{\nu}}\left(\f{z^2}{2s}\right)-I_{\ti{\nu}}\left(\f{z^2}{2s}\right)\right) \no
&&= -\f{V_{d-1}}{12} \int^\infty_\ep \f{dz}{z}\int^\infty_{0}
\f{ds}{2s}\cdot (4\pi s)^{-\f{d-1}{2}}e^{-\f{z^2}{2s}}\int^{\nu^2}_0 d\ti{\nu}^2
\f{2}{\pi}\sin(\pi\ti{\nu})K_{\ti{\nu}}\left(\f{z^2}{2s}\right)\no
&&= -\f{V_{d-1}}{12 (4\pi)^{\f{d-1}{2}}}
     \int_{\ep}^{\infty} \f{dz}{z} \int_{\d^{2}}^{\infty} \f{ds}{s}
     s^{- \f{d-1}{2}} e^{- \f{z^2}{2s}}
     \int_{0}^{\nu^2} d \tilde{\nu}^2 \f{K_{\tilde{\nu}}(\f{z^2}{2s})}
     {\Gamma(\tilde{\nu}) \Gamma(1-\tilde{\nu})} \\
  && = -\f{V_{d-1}\s{\pi}}{12 (4\pi)^{\f{d-1}{2}} (d-1) \Gamma(\f d 2)\ep^{d-1}} \no
     && \hspace{50pt} \times \int_{0}^{\nu^2} d \tilde{\nu}^2
     \f{\Gamma(\f{d-1}{2}+\tilde{\nu}) \Gamma(\f{d-1}{2}-\tilde{\nu})}{\Gamma(\tilde{\nu}) \Gamma(1-\tilde{\nu})} \\
  && = -\f{V_{d-1}\s{\pi}}{12 (4\pi)^{\f{d-1}{2}} (d-1) \Gamma(\f d 2)\ep^{d-1}} \no
     && \hspace{50pt} \times \int_{0}^{\nu^2} d \tilde{\nu}^2
     \begin{cases}
      \prod^{\f{d-2}{2}}_{n=1}[(\f{d-1}{2} - n)^2 - \tilde{\nu}^2] \tan(\pi \tilde{\nu}) & \text{($d \in$even)}\\
      \prod^{\f{d-3}{2}}_{n=1}[(\f{d-1}{2} - n)^2 - \tilde{\nu}^2] \tilde{\nu} & \text{($d \in$odd)}
     \end{cases} \label{resf}
\ea
When $d$ is an odd integer and $d\geq 3$, the above result (\ref{resf}) agrees with (\ref{ee odd}). Even though our analysis can not be applied to $d=1,2$, the final result reproduces (\ref{eedd1pp}) and (\ref{eed2pp}).\footnote{We would also like to mention that only when $d=2$, the difference (\ref{resf}) gets divergent for
$\nu\geq 1/2$. This might suggest there is a phase transition for $\nu\geq 1/2$ in our replicated theory, which deserves a future study. For other values of $d$, the difference is always finite.}
At the same time it offers results for general even $d$.

\section{Entanglement Entropy and Double Trace Deformations}

Here we combine the bulk entanglement entropy and the quantum back reaction effects to obtain the one-loop holographic entanglement entropy (\ref{loopee}). We are interested in the difference between the entropy for the Neumann boundary condition and that for the Dirichlet one, which is denoted as $\Delta S^{1-loop}_A=S^{1-loop}_A(-\nu)-S^{1-loop}_A(\nu)$. Thus we subtract all bulk UV divergences in a common way and this naturally takes into account counter term contributions $S_{c.t.}$ or equally renormalizes the Newton constant.\footnote{Refer to e.g. the review
\cite{BHS} and the references therein for a similar analysis which relates the entanglement entropy to the quantum corrections of black hole entropy.}

\subsection{Bulk Entanglement Entropy}

We already computed the bulk entanglement entropy $S^{bulk}_A$ in (\ref{ee odd}) and (\ref{resf}). The result is summarized as
\be
\Delta S^{bulk}_A= -\f{V_{d-1}}{(d-1) \ep^{d-1}}\cdot\f{\s{\pi}}{6 (4\pi)^{\f{d-1}{2}} \Gamma(\f d 2)}  \int_{0}^{\nu} d \tilde{\nu}\cdot\ti{\nu}\cdot
     \f{\Gamma(\f{d-1}{2}+\tilde{\nu}) \Gamma(\f{d-1}{2}-\tilde{\nu})}{\Gamma(\tilde{\nu}) \Gamma(1-\tilde{\nu})} .
\ee

For $d=1$ (AdS$_3$), $d=3$ (AdS$_5$), $d=5$ (AdS$_7$) and $d=7$ (AdS$_9$), we explicitly find
\ba
&&\Delta S^{bulk(d=1)}_A=\f{\nu}{6}\log(\ep^{-1}), \ \ \ \ \Delta S^{bulk(d=3)}_A=\f{V_{2}}{2\ep^2}\cdot \left(-\f{\nu^3}{36\pi}\right), \label{qq}\\
&&\Delta S^{bulk(d=5)}_A=\f{V_{4}}{4\ep^4}\cdot \left(-\f{1}{72\pi^2}\right)\cdot\left(\f{\nu^3}{3}-\f{\nu^5}{5}\right).\no
&&\Delta S^{bulk(d=7)}_A=\f{V_{6}}{6\ep^6}\cdot \left(-\f{1}{720\pi^3}\right)\cdot\left(\f{4}{3}\nu^3-\nu^5+\f{\nu^7}{7}\right).
\ea

\subsection{Minimal Area Corrections}

To compute the shift of minimal area surface i.e. the term $\f{\delta \mbox{Area}}{4G_N}$
in (\ref{loopee}), we need to solve the back reaction of the metric due to
quantum corrections to the bulk energy stress tensor $T_{\mu\nu}$.
 The Euclidean gravity action coupled to a massive scalar is
 \ba
&& I_G=-\f{1}{16\pi G_N}\int \s{g}({\ca R}-2\Lambda)+\int \s{g}L_{scalar},\no
&& L_{scalar}=\f{1}{2}(\de_\mu\phi)^2+\f{1}{2}m^2\phi^2.
 \ea

Using $\delta \s{g}=\f{1}{2}\s{g}\delta g_{\mu\nu}g^{\mu\nu}=-\f{1}{2}\s{g}\delta g^{\mu\nu}g_{\mu\nu}$, we have
\ba
\delta I_G=-\f{1}{16\pi G_N}\int \s{g}\delta g^{\mu\nu}({\ca R}_{\mu\nu}-\f{1}{2}{\ca R}g_{\mu\nu}+\Lambda g_{\mu\nu})
+\f{1}{2}\int \s{g}\delta g^{\mu\nu}T_{\mu\nu},
\ea
where $T_{\mu\nu}=\de_\mu\phi\de_\nu\phi-\f{1}{2}g_{\mu\nu}((\de\phi)^2+m^2\phi^2)$.
This leads to the Einstein equation
\be
{\ca R}_{\mu\nu}-\f{1}{2}{\ca R}g_{\mu\nu}+\Lambda g_{\mu\nu}=8\pi G_N T_{\mu\nu}.
\ee

For a pure AdS$_{d+2}$ with the AdS radius $R$ we have
\be
{\ca R}=-\f{(d+1)(d+2)}{R^2},\ \ \ {\ca R}_{\mu\nu}=-\f{d+1}{R^2}g_{\mu\nu},\ \ \
\Lambda=-\f{d(d+1)}{2R^2}.
\ee
Note also that the gravity action for this classical solution of AdS$_{d+2}$ is evaluated as
\be
-\f{1}{16\pi G_N}\int \s{g}({\ca R}-2\Lambda)=\f{d+1}{8\pi G_N R^2}\int_{AdS} \s{g}. \label{cac}
\ee

If the quantum expectation value of $T_{\mu\nu}$ is given by $\la T_{\mu\nu}\lb=\lambda g_{\mu\nu}$,
the radius $\ti{R}$ of the corrected background is given by
\be
\f{1}{\ti{R}^2}=\f{1}{R^2}+\f{16\pi G_N}{d(d+1)}\lambda.
\ee
By assuming $G_N\lambda$ is very small, we have
\be
\delta R=\ti{R}-R=-\f{8\pi G_N R^3}{d(d+1)}\lambda.
\ee

The (renormalized) quantum expectation values of $T_{\mu\nu}$ for the Dirichlet boundary condition was computed
in \cite{Cal}. For $d=1,3,5,7$ (AdS$_{3,5,7,9}$), we explicitly find
\ba
&& \la T_{\mu\nu}\lb_{d=1}(\nu)=-\f{(1-\nu^2)\nu}{12\pi R^3}g_{\mu\nu}, \ \ \ \ \la T_{\mu\nu}\lb_{d=3}(\nu)=-\f{(4-\nu^2)(1-\nu^2)\nu}{120\pi^2 R^5}g_{\mu\nu}, \no
&& \la T_{\mu\nu}\lb_{d=5}(\nu)=-\f{(9-\nu^2)(4-\nu^2)(1-\nu^2)\nu}{1680\pi^3 R^7}g_{\mu\nu}, \no
&& \la T_{\mu\nu}\lb_{d=7}(\nu)=-\f{(16-\nu^2)(9-\nu^2)(4-\nu^2)(1-\nu^2)\nu}{30240\pi^4 R^9}g_{\mu\nu}.
\ea
Thus the difference (Neumann$-$Dirichlet) between the two boundary condition are given by
\ba
\Delta\la T_{\mu\nu}\lb_{d}=-2\la T_{\mu\nu}\lb_{d}(\nu),
\ea
for any $d$.

The area of minimal surface after the radius shift $R\to \ti{R}$ due to the quantum correction is given by $\f{V_{d-1}}{(d-1)\ep^{d-1}}\cdot \f{\ti{R}^d}{4G_N}$. Therefore we find the change of minimal area surface is estimated as
\be
\f{\delta \mbox{Area}}{4G_N}=\f{V_{d-1}}{(d-1)\ep^{d-1}}\cdot \f{d R^{d-1}\cdot \delta R}{4G_N}
=-\f{2\pi V_{d-1}}{(d^2-1)\ep^{d-1}}R^{d+2}\lambda.
\ee

For $d=1,3,5,7$, this leads to
\ba
&& \Delta \left(\f{\delta \mbox{Area}}{4G_N}\right)_{d=1}=\log(\ep^{-1})\cdot  \left(\f{\nu^3-\nu}{6}\right),\ \ \  \Delta \left(\f{\delta \mbox{Area}}{4G_N}\right)_{d=3}=\f{V_{2}}{2\ep^{2}}\cdot
\left(-\f{\nu(4-\nu^2)(1-\nu^2)}{120\pi}\right),\no
&& \Delta \left(\f{\delta \mbox{Area}}{4G_N}\right)_{d=5}=\f{V_{4}}{4\ep^{4}}\cdot
\left(-\f{\nu(9-\nu^2)(4-\nu^2)(1-\nu^2)}{2520\pi^2}\right),\no
&& \Delta \left(\f{\delta \mbox{Area}}{4G_N}\right)_{d=7}=\f{V_{6}}{6\ep^{6}}\cdot
\left(-\f{\nu(16-\nu^2)(9-\nu^2)(4-\nu^2)(1-\nu^2)}{60480\pi^3}\right), \label{qqq}
\ea

\subsection{Total Contributions}

For $d=1,3,5,7$, we sum up (\ref{qq}) and (\ref{qqq}) and get
\ba
&&\Delta S^{1-loop(d=1)}_A=\f{\nu^3}{6}\log(\ep^{-1}). \label{loopa}\\
&&\Delta S^{1-loop(d=3)}_A=\f{V_2}{2\pi\ep^2}\cdot\left(
-\f{\nu^5}{120}+\f{\nu^3}{72}-\f{\nu}{30}\right),\label{loopb} \\
&&\Delta S^{1-loop(d=5)}_A=\f{V_4}{4\pi^2\ep^4}\cdot\left(
\f{1}{2520}\nu^7-\f{1}{360}\nu^5+\f{2}{135}\nu^3-\f{1}{70}\nu\right), \label{loopc}\\
&&\Delta S^{1-loop(d=7)}_A=\f{V_6}{6\pi^3\ep^6}\cdot\left(
-\f{1}{60480}\nu^9+\f{1}{3360}\nu^7-\f{1}{320}\nu^5+\f{59}{5040}\nu^3-\f{1}{105}\nu\right). \label{loopd}
\ea
Notice that only for $d=1$, the change of entanglement entropy is positive, while it is negative
in higher dimensions. Also all of them vanish at $\nu=0$ because the Neumann and Dirichlet boundary condition become equivalent as can be seen from (\ref{dome}).

\subsection{Comparison with Central Charge}

In \cite{GM}, the shift of vacuum energy in AdS between the two boundary conditions has been computed:
\be
\Delta V=\f{(-1)^{\f{d-1}{2}}}{2^{\f{d+1}{2}}\cdot (d!!)\cdot \pi^{\f{d+1}{2}}R^{d+2}}
\int^\nu_0 dx \prod_{j=0}^{\f{d-1}{2}}(x^2-j^2),
\ee
which is always positive.
For $d=1,3,5,7$, we explicitly find
\ba
&& \Delta V_{d=1}=\f{\nu^3}{6\pi R^3},\label{kk} \\
&& \Delta V_{d=3}=\f{1}{12\pi^2 R^5}\left(-\f{\nu^5}{5}+\f{\nu^3}{3}\right), \no
&& \Delta V_{d=5}=\f{1}{120\pi^3 R^7}\left(\f{\nu^7}{7}-\nu^5+\f{4}{3}\nu^3\right), \no
&& \Delta V_{d=7}=\f{1}{1680\pi^4 R^9}\left(-\f{\nu^9}{9}+2\nu^7-\f{49}{5}\nu^5+12\nu^3\right).
\ea

Since the vacuum energy $V$ contributes to the gravity action as $S_{1-loop}=V\int_{AdS} \s{g}$,
we can obtain the shift of central charge $\Delta c$  by comparing $V$ with the action (\ref{cac}) \cite{GM}:
\be
\f{\Delta c}{c}=\Delta V\cdot \f{8\pi G_N R^2}{d+1}. \label{rati}
\ee
Note that this central charge is the special one (often called $a$ in four dimensions)
which is expected to decrease monotonically under the RG flow.

For $d=1$, this predicts the following correction compared with the tree level result $S^{tree}_A=\f{c}{6}\log \ep^{-1}$:
\be
\Delta S^{1-loop(d=1)}_A=\left(\f{\Delta c}{c}\right)\cdot S^{tree(d=1)}_A=\f{\nu^3}{6}\log (\ep^{-1}),
\ee
where we employed (\ref{kk}) as expected. This agrees with our one-loop holographic entanglement entropy (\ref{loopa}).

For $d=3,5,7$, the leading divergent pieces of entanglement entropy (i.e. area law term) are not proportional to the central charge in general. If we naively compute the analogue of
(\ref{rati}) for these, we obtain
\ba
&& \left(\f{\Delta c}{c}\right)\cdot S^{tree(d=3)}_A=\f{V_2}{2\pi\ep^2}\cdot\left(-\f{\nu^5}{120}+\f{\nu^3}{72}
\right),\no
&& \left(\f{\Delta c}{c}\right)\cdot S^{tree(d=5)}_A=\f{V_4}{4\pi^2\ep^4}\cdot \left(\f{\nu^7}{2520}-\f{\nu^5}{360}+\f{\n^3}{270}\right),\no
&& \left(\f{\Delta c}{c}\right)\cdot S^{tree(d=7)}_A=\f{V_6}{6\pi^3\ep^6}\cdot
\left(-\f{\nu^9}{60480}+\f{\nu^7}{3360}-\f{7}{4800}\nu^5+\f{\n^3}{560}\right).\no
\ea
Indeed, they deviate from the one-loop holographic entanglement entropy computed in
(\ref{loopb}), (\ref{loopc}) and (\ref{loopd}). It is intriguing to observe that for all of these, the first two coefficients (i.e. $\nu^{d+2}$ and $\nu^{d}$) agree with each other after non-trivial
cancellations, which might be an interesting future problem to explain.

\section{Conclusions}

In the first part of this paper, we computed the entanglement entropy of a free massive scalar field analytically in the bulk AdS space. We considered two boundary conditions (i.e. Dirichlet and Neumann) of the scalar field at the AdS boundary. In both cases we found that the bulk UV divergent terms take the same expression and that the leading divergent term follows the area law. We showed that the finite terms depend on the boundary conditions and we computed the difference between these two cases. We computed this finite entropy difference from two methods: a direct partition function
computation and a calculation based on Green functions, both of which lead to the same result.

Next, by adding the quantum back-reaction of minimal surface area to the bulk entanglement entropy, we found the one-loop corrections to the holographic entanglement entropy. In the case of
AdS$_3/$CFT$_2$, we confirmed that our result reproduces the known expression in two dimensional CFTs, which is proportional to the central charge. In this analysis, non-trivial cancelations of terms in the two contributions play an important role. Note that in this example, the entanglement entropy is decreased under the RG flow and this is consistent with the c-theorem \cite{Za,CaHu}.

Moreover, we computed the one-loop corrections to the entanglement entropy in higher dimensions.  We found that the difference becomes negative and this means that the entanglement entropy is not monotonically decreasing under the RG flow as opposed to the AdS$_3$ case. The results are no longer proportional to the central charge and the relevant results in CFTs have not been known so far. It will be an intriguing future problem to reproduce the same result from field theory computations.

Finally we would like to comment that in our analysis, we computed the coefficient of area law divergences of entanglement entropy in $d+1$ dimensional CFTs (i.e. the UV and IR fixed point of the double trace deformation). Except $d=1$, one may wonder if the coefficient is universal as it depends on the choice of the UV cut off $\ep$ of CFTs. However, since we are performing the holographic computation, the geometric cut off $z>\ep$ is universal. Indeed we can explicitly confirm from our result in section 3.1. that the bulk UV divergent terms $O(\delta^{-d})$ in the difference of bulk entanglement entropy $\Delta S^{bulk}_{A}(-\nu)-\Delta S^{bulk}_{A}(\nu)$, does cancel only if we assume that the same CFT UV cut off $z>\ep$ is introduced in the two holographic CFTs.

We comment on the case that the subsystem is a sphere $S^{d-1}$ in $d+1$ dimensional CFTs.\footnote{We would like to thank Igor Klebanov for pointing out this issue.} The entanglement entropy of a sphere $S^{d-1}$ in CFTs can be mapped to the free energy on the Euclidean sphere $S^{d+1}$ \cite{CaHuMy}. The double-trace deformation on $S^{d+1}$ was studied in \cite{GK,DiDo,KlPuSa}(see also further related works \cite{GiKl, GiKlPuSaTa}). So, when the subsytem is a sphere, it would be possible to calculate how much the entanglement entropy changes under a double-trace deformation in CFTs and compare it with its AdS dual.

\subsection*{Acknowledgements}

We would like to thank Shunji Matsuura, Yu Nakayama, Tokiro Numasawa, Kento Watanabe and Satoshi Yamaguchi for useful comments and especially Thomas Faulkner and Aitor Lewkowycz for reading our draft and valuable communications. We also would like to thank Igor R. Klebanov for useful comments on the case that the subsystem is a sphere. TT is also grateful to Juan Maldacena for useful conversations on this work.  TT is supported by JSPS Grant-in-Aid for Scientific Research (B) No.25287058. TT is also supported by World Premier International Research Center Initiative (WPI Initiative) from the Japan Ministry of Education, Culture, Sports, Science and Technology (MEXT). NS is supported by a JSPS fellowship.
\begin{appendix}


\section{Direct Replica Method Calculations}


We can obtain the Green function  in the $n$-sheeted $AdS_{d+2}$
in the same way as
the Green function in the $n$-sheeted flat space in \cite{Kabat,CC}.
The propagator in the $n$-sheeted $AdS_{d+2}$ for $z<z'$ is given by
\begin{equation}
\begin{split}
&\bra{r,\theta,x,z}\dfrac{1}{\Box-m^2}\ket{r',\theta',x',z'}(\nu) \\
&= \dfrac{-(z z')^{\frac{d+1}{2}}}{R^{d}} \dfrac{1}{2\pi n}
\sum_{l=0}^\infty
d_l \int \dfrac{d^{d-1}k}{(2\pi)^{d-1}}
 \int_0^\infty d\beta \beta
J_{l/n}(\beta r) J_{l/n}(\beta r') \\
&\times \cos \left( \dfrac{l}{n}(\theta-\theta') \right)
e^{ik\cdot(x-x')} I_{\nu}(\sqrt{\beta^2+k^2}z)
 K_{\nu}(\sqrt{\beta^2+k^2}z'),
\end{split}
\end{equation}
where $x_0=r\cos \theta, x_1=r\sin \theta, x=(x_2,x_3,\cdots,x_d)$
and $d_0=1,d_{l>0}=2$.
Thus we obtain
\begin{equation}
\begin{split}
&\dfrac{\partial}{\partial \nu^2}[ \ln Z_n (-\nu)-\ln Z_n (\nu)] \\
&=-\dfrac{1}{2} V_{d-1} \dfrac{2}{\pi}\sin \nu \pi
\int_0^\infty dr r \int_\epsilon^\infty \dfrac{dz}{z} \sum_{l=0}^\infty
d_l \int \dfrac{d^{d-1}k}{(2\pi)^{d-1}} \\
&\times \int_0^\infty d\beta \beta
(J_{l/n}(\beta r))^2 (K_{\nu}(\sqrt{\beta^2+k^2}z))^2  \\
&=-\dfrac{1}{2} V_{d-1} \dfrac{2}{\pi}\sin \nu \pi
\int_0^\infty dr r \int_\epsilon^\infty \dfrac{dz}{z} \sum_{l=0}^\infty
d_l \int \dfrac{d^{d-1}k}{(2\pi)^{d-1}} \\
&\times \int_0^\infty d\beta \beta
(J_{l/n}(\beta r))^2
 \int_0^\infty \dfrac{ds}{2s}K_{\nu}\left(\dfrac{z^2}{2s}\right)
\exp \left[-s(\beta^2+k^2)-\dfrac{z^2}{2s} \right]  \\
&=-\dfrac{1}{2} V_{d-1} \dfrac{2}{\pi}\sin \nu \pi
\int_0^\infty dr r \int_\epsilon^\infty \dfrac{dz}{z} \sum_{l=0}^\infty
d_l \int \dfrac{d^{d-1}k}{(2\pi)^{d-1}} \\
&\times
 \int_0^\infty \dfrac{ds}{2s}K_{\nu}\left(\dfrac{z^2}{2s}\right)
\exp \left[-s k^2-\dfrac{z^2}{2s} \right]
\dfrac{1}{2s} \exp \left[-\dfrac{r^2}{2s}\right] I_{l/n}\left(\dfrac{r^2}{2s}\right) \\
&=-\dfrac{1}{2} V_{d-1} \dfrac{2}{\pi}\sin \nu \pi
 \int_\epsilon^\infty \dfrac{dz}{z} \sum_{l=0}^\infty
d_l \int \dfrac{d^{d-1}k}{(2\pi)^{d-1}} 
 \int_0^\infty \dfrac{ds}{2s}K_{\nu}\left(\dfrac{z^2}{2s}\right)
\exp \left[-s k^2-\dfrac{z^2}{2s} \right]
\left(\dfrac{-l}{2n} \right) \\
&=-\dfrac{1}{2} V_{d-1} \dfrac{2}{\pi}\sin \nu \pi
\int_\epsilon^\infty \dfrac{dz}{z}
\dfrac{1}{24n} \dfrac{1}{(4\pi)^{\frac{d-1}{2}}}
 \int_0^\infty \dfrac{ds}{s^{\frac{d+1}{2}}}K_{\nu}\left(\dfrac{z^2}{2s}\right)
\exp \left[-\dfrac{z^2}{2s} \right] \\
&=-\dfrac{1}{2} V_{d-1} \dfrac{2}{\pi}\sin \nu \pi
\dfrac{1}{24n} \dfrac{1}{(4\pi)^{\frac{d-1}{2}}}
\sqrt{\pi} \dfrac{\Gamma(\rho_d+\nu) \Gamma(\rho_d-\nu)}{\Gamma(d/2)}
\int_\epsilon^\infty \dfrac{dz}{z^d} ,
\end{split}  \label{voll}
\end{equation}
where $\rho_d=\frac{1}{2}(d-1)$ and
we used $\sum_{l=0}^\infty d_l l=2\zeta(-1)=-\frac{1}{6} $.
Finally, we obtain
($S_A=-\frac{\partial}{\partial n}[\ln Z_n -n\ln Z_1]|_{n=1}$)
\begin{equation}
\begin{split}
&\dfrac{\partial}{\partial \nu^2}[ S_A (-\nu)-S_A (\nu)] \\
&=-\dfrac{1}{2} V_{d-1} \dfrac{2}{\pi}\sin \nu \pi
\dfrac{1}{12} \dfrac{1}{(4\pi)^{\frac{d-1}{2}}}
\sqrt{\pi} \dfrac{\Gamma(\rho_d+\nu) \Gamma(\rho_d-\nu)}{\Gamma(d/2)}
\int_\epsilon^\infty \dfrac{dz}{z^d} .
\end{split}  \label{vol}
\end{equation}
This result is same as (\ref{resf}) obtained by the orbifold method.

\section{Green function}
In the same way as the partition function, we obtain the Green function as
\begin{equation}
\begin{split}
\bra{x} \frac{1}{\Box-m^2} \ket{x'}
& =  \frac{-(z z')^{\frac{d+1}{2}}}{R^d}  \int \frac{d^{d+1}p}{(2\pi)^{d+1}}
e^{i\vec{p} \cdot (\vec{x}-\vec{x'}) }
 \int_0^{\infty} d \alpha \mu (\alpha) \frac{1}{\alpha^2+\nu^2}  K_{i\alpha}(pz)  K_{i\alpha}(pz')    \\
&= \frac{-(z z')^{\frac{d+1}{2}}}{R^d}  \int \frac{d^{d+1}p}{(2\pi)^{d+1}}
e^{i\vec{p} \cdot (\vec{x}-\vec{x'}) }
\frac{1}{2i\pi} \int_{-\infty}^{\infty} d \alpha \frac{2\alpha}{\alpha^2 +\nu^2} I_{-i\alpha} (pz)
K_{i\alpha}(pz'),
\end{split}  \label{Green}
\end{equation}
where we have used $K_{i\alpha} =\frac{\pi}{2} \frac{I_{-i\alpha}-I_{i\alpha}}{i \sinh \alpha \pi}$.
For $z<z'$, $ I_{-i\alpha} (pz) K_{i\alpha}(pz')$ is expressed as
\begin{equation}
\begin{split}
 I_{-i\alpha} (pz) K_{i\alpha}(pz')
&= \int_0^{\infty} \dfrac{ds}{2s} e^{-sp^2} e^{-\frac{z^2+z'^2}{4s}}
I_{-i\alpha} \left(\dfrac{z z'}{2s}  \right) \\
&= \int_0^{\infty} \dfrac{ds}{2s} e^{-sp^2} e^{-\frac{z^2+z'^2}{4s}}
\dfrac{1}{2i\pi} \int_{\infty -i\pi}^{\infty+i\pi}
\exp \left( \dfrac{z z'}{2s} \cosh t +i\alpha t  \right) dt .
\end{split}
\end{equation}
By using this integral representation, we perform the $\alpha$ integral and obtain
\begin{equation}
\begin{split}
&\frac{1}{2i\pi} \int_{-\infty}^{\infty} d \alpha \frac{2\alpha}{\alpha^2 +\nu^2} I_{-i\alpha} (pz)
K_{i\alpha}(pz') \\
&=\int_0^{\infty} \dfrac{ds}{2s} e^{-sp^2} e^{-\frac{z^2+z'^2}{4s}}
\dfrac{1}{2i\pi} \int_{\infty -i\pi}^{\infty+i\pi}
\exp \left( \dfrac{z z'}{2s} \cosh t -\nu t  \right) dt  \\
&= I_{\nu} (pz) K_{\nu}(pz') .
\end{split} \label{Green integral}
\end{equation}
By substituting (\ref{Green integral}) into (\ref{Green}),
we obtain the Green function as
\begin{equation}
\begin{split}
\bra{x} \frac{1}{\Box-m^2} \ket{x'}
& =  \frac{-(z z')^{\frac{d+1}{2}}}{R^d}  \int \frac{d^{d+1}p}{(2\pi)^{d+1}}
e^{i\vec{p} \cdot (\vec{x}-\vec{x'}) }
  I_{\nu}(pz)  K_{\nu}(pz') .   \\
\end{split}
\end{equation}
This is the Green function for the Dirichlet boundary condition (see e.g. \cite{GM}).

\section{A derivation of the orthogonal relation (\ref{orthogonal})}\label{ap:or}

For $a,b,c \in \mbox{R}$ and $c >0$, we define
\begin{equation}
\begin{split}
&A(a,b,c) \equiv \int_0^{\infty} dz \dfrac{1}{z^{1-c}} K_{i a}(z) K_{i b}(z)  \\
&=\frac{2^{-3+c}}{ \Gamma(c)} \Gamma \left(\frac{i}{2} (a-b-ic) \right)
\Gamma \left(\frac{i}{2} (a+b-ic) \right)
\Gamma \left(\frac{-i}{2} (a-b+ic) \right)
\Gamma \left(\frac{-i}{2} (a+b+ic) \right)
\end{split}
\end{equation}
We take the limit $c\to 0$.
For $a \neq b$ we obtain $\lim_{c\to 0}A(a,b,c) =0 $.
Thus, when $c\to 0$, we obtain
\begin{equation}
\begin{split}
A(a,b,c)
&\sim \frac{2^{-3+c}}{ \Gamma(c)} \Gamma \left(i a \right)
\Gamma \left(-i a \right)
\Gamma \left(\frac{-i}{2} (a-b+ic) \right)
\Gamma \left(\frac{i}{2} (a-b-ic) \right)   \\
&=\frac{\pi^2}{2a \sinh \pi a } \frac{2}{\pi} \frac{2^{-3+c}}{ \Gamma(c)}
\Gamma \left(\frac{-i}{2} (a-b+ic) \right)
\Gamma \left(\frac{i}{2} (a-b-ic) \right)   \\
&\sim \frac{\pi^2}{2a \sinh \pi a } \frac{2}{\pi} 2^{-3+c}
\frac{4c}{(a-b)^2+c^2 }   \\
&\to \frac{\pi^2}{2a \sinh \pi a } \delta(a-b) .  \\
\end{split}
\end{equation}

\end{appendix}



\begin{thebibliography}{}




\bibitem{Ma}
  J.~M.~Maldacena,
  Adv.\ Theor.\ Math.\ Phys.\  {\bf 2} (1998) 231
  [Int.\ J.\ Theor.\ Phys.\  {\bf 38} (1999) 1113]
  [arXiv:hep-th/9711200];

\bibitem{GKP}
 S.~S.~Gubser, I.~R.~Klebanov and A.~M.~Polyakov,
 Phys.\ Lett.\  B {\bf 428}, 105 (1998)
  [arXiv:hep-th/9802109].

\bibitem{Wa}
E.~Witten,
  Adv.\ Theor.\ Math.\ Phys.\  {\bf 2}, 253 (1998)
  [arXiv:hep-th/9802150].

\bibitem{RT}
 S.~Ryu and T.~Takayanagi,
 Phys.\ Rev.\ Lett.\  {\bf 96} (2006) 181602  [hep-th/0603001];
  JHEP {\bf 0608}, 045 (2006)
  [hep-th/0605073];
V.~E.~Hubeny, M.~Rangamani and T.~Takayanagi,
``A Covariant holographic entanglement entropy proposal,''  JHEP {\bf 0707} (2007) 062  [arXiv:0705.0016 [hep-th]].  


\bibitem{Review}
T.~Nishioka, S.~Ryu and T.~Takayanagi,
 ``Holographic Entanglement Entropy: An Overview,''
  J.\ Phys.\ A  {\bf 42} (2009) 504008;
 T.~Takayanagi,
  ``Entanglement Entropy from a Holographic Viewpoint,''
  Class.\ Quant.\ Grav.\  {\bf 29} (2012) 153001  [arXiv:1204.2450 [gr-qc]].


\bibitem{BDHM}
  T.~Barrella, X.~Dong, S.~A.~Hartnoll and V.~L.~Martin,
  ``Holographic entanglement beyond classical gravity,''  JHEP {\bf 1309} (2013) 109  [arXiv:1306.4682 [hep-th]].  



\bibitem{FLM}
  T.~Faulkner, A.~Lewkowycz and J.~Maldacena,
  ``Quantum corrections to holographic entanglement entropy,''  JHEP {\bf 1311} (2013) 074  [arXiv:1307.2892 [hep-th]].  


\bibitem{He}
  M.~Headrick,
  ``Entanglement Renyi entropies in holographic theories,''  Phys.\ Rev.\ D {\bf 82} (2010) 126010  [arXiv:1006.0047 [hep-th]].  

\bibitem{Ha}
  T.~Hartman,
  ``Entanglement Entropy at Large Central Charge,''  arXiv:1303.6955 [hep-th].  


\bibitem{Fa}
  T.~Faulkner,
  ``The Entanglement Renyi Entropies of Disjoint Intervals in AdS/CFT,''  arXiv:1303.7221 [hep-th].  


\bibitem{Shiba}
  N.~Shiba,
  ``Entanglement Entropy of Two Spheres,''  JHEP {\bf 1207} (2012) 100  [arXiv:1201.4865 [hep-th]].  

\bibitem{Cardy}
  J.~Cardy,
  ``Some results on the mutual information of disjoint regions in higher dimensions,''  J.\ Phys.\ A {\bf 46} (2013) 285402  [arXiv:1304.7985 [hep-th]].  

\bibitem{AF}
C.~Agon and T.~Faulkner, ``Quantum Corrections to Holographic Mutual Information'', [arXiv:1511.07462 [hep-th]] 

\bibitem{Bin}
B.~Chen and J.~J.~Zhang,
  ``On short interval expansion of Rényi entropy,''  JHEP {\bf 1311} (2013) 164  [arXiv:1309.5453 [hep-th]].  

\bibitem{DD}
S.~Datta and J.~R.~David,
  ``Rényi entropies of free bosons on the torus and holography,''  JHEP {\bf 1404} (2014) 081  [arXiv:1311.1218 [hep-th]].  

\bibitem{Pe}
  E.~Perlmutter,
  ``Comments on Renyi entropy in AdS$_3$/CFT$_2$,''  JHEP {\bf 1405} (2014) 052  [arXiv:1312.5740 [hep-th]].  

\bibitem{CSZ}
  B.~Chen, F.~y.~Song and J.~j.~Zhang,
  ``Holographic Renyi entropy in AdS$_3$/LCFT$_2$ correspondence,''  JHEP {\bf 1403} (2014) 137  [arXiv:1401.0261 [hep-th]].  


\bibitem{BeMa}
  M.~Beccaria and G.~Macorini,
  ``On the next-to-leading holographic entanglement entropy in $AdS_{3}/CFT_{2}$,''  JHEP {\bf 1404} (2014) 045  [arXiv:1402.0659 [hep-th]].  




\bibitem{ShibaE}
  N.~Shiba,
  ``Entanglement Entropy of Disjoint Regions in Excited States : An Operator Method,''  JHEP {\bf 1412} (2014) 152  [arXiv:1408.0637 [hep-th]].  

\bibitem{NaNi}
  Y.~Nakaguchi and T.~Nishioka,
  ``Entanglement Entropy of Annulus in Three Dimensions,''  JHEP {\bf 1504} (2015) 072  [arXiv:1501.01293 [hep-th]].  



\bibitem{HMPZ}
  M.~Headrick, A.~Maloney, E.~Perlmutter and I.~G.~Zadeh,
  ``Rényi entropies, the analytic bootstrap, and 3D quantum gravity at higher genus,''  JHEP {\bf 1507} (2015) 059  [arXiv:1503.07111 [hep-th]].  



\bibitem{SHS}
  B.~Swingle, L.~Huijse and S.~Sachdev,
  ``Entanglement entropy of compressible holographic matter: loop corrections from bulk fermions,''  Phys.\ Rev.\ B {\bf 90} (2014) 4,  045107  [arXiv:1308.3234 [hep-th]].  


\bibitem{SwRa}
  B.~Swingle and M.~Van Raamsdonk,
  ``Universality of Gravity from Entanglement,''  arXiv:1405.2933 [hep-th].  


\bibitem{EW}
  N.~Engelhardt and A.~C.~Wall,
  ``Quantum Extremal Surfaces: Holographic Entanglement Entropy beyond the Classical Regime,''  JHEP {\bf 1501} (2015) 073  [arXiv:1408.3203 [hep-th]].  

\bibitem{Le}
  S.~Leichenauer,
  ``Thermal Corrections to Entanglement Entropy from Holography,''  JHEP {\bf 1509} (2015) 014  [arXiv:1502.07348 [hep-th]].


\bibitem{Kelly:2015mna}
  W.~R.~Kelly, K.~Kuns and D.~Marolf,
  ``’t Hooft suppression and holographic entropy,''  JHEP {\bf 1510} (2015) 059  [arXiv:1507.03654 [hep-th]].  



\bibitem{Wit}
  E.~Witten,
  ``Multitrace operators, boundary conditions, and AdS / CFT correspondence,''  hep-th/0112258.  



\bibitem{GM}
  S.~S.~Gubser and I.~Mitra,
  ``Double trace operators and one loop vacuum energy in AdS / CFT,''  Phys.\ Rev.\ D {\bf 67} (2003) 064018  [hep-th/0210093].  

\bibitem{GK}
  S.~S.~Gubser and I.~R.~Klebanov,
  ``A Universal result on central charges in the presence of double trace deformations,''  Nucl.\ Phys.\ B {\bf 656} (2003) 23  [hep-th/0212138].  



\bibitem{BKLS}
  L.~Bombelli, R.~K.~Koul, J.~Lee and R.~D.~Sorkin,
  ``A Quantum Source of Entropy for Black Holes,''
  Phys.\ Rev.\ D {\bf 34} (1986) 373.

\bibitem{Sr}
   M.~Srednicki,
   ``Entropy and area,''
   Phys.\ Rev.\ Lett.\  {\bf 71} (1993) 666  [hep-th/9303048].


\bibitem{LM}
  A.~Lewkowycz and J.~Maldacena,
  ``Generalized gravitational entropy,''  JHEP {\bf 1308} (2013) 090  [arXiv:1304.4926 [hep-th]].  



\bibitem{KW}
  I.~R.~Klebanov and E.~Witten,
  ``AdS / CFT correspondence and symmetry breaking,''  Nucl.\ Phys.\ B {\bf 556} (1999) 89  [hep-th/9905104].  



\bibitem{HLW}
  C.~Holzhey, F.~Larsen and F.~Wilczek,
  ``Geometric and renormalized entropy in conformal field theory,''  Nucl.\ Phys.\ B {\bf 424} (1994) 443  [hep-th/9403108].  

\bibitem{CC}
 P.~Calabrese and J.~L.~Cardy,
  ``Entanglement entropy and quantum field theory,''
  J.\ Stat.\ Mech.\  {\bf 0406}, P06002 (2004)
  [hep-th/0405152].


\bibitem{BrHe}
  J.~D.~Brown and M.~Henneaux,
  ``Central Charges in the Canonical Realization of Asymptotic Symmetries: An Example from Three-Dimensional Gravity,''  Commun.\ Math.\ Phys.\  {\bf 104} (1986) 207.  



\bibitem{NiTa}
  T.~Nishioka and T.~Takayanagi,
  ``AdS Bubbles, Entropy and Closed String Tachyons,''  JHEP {\bf 0701} (2007) 090  [hep-th/0611035].  

\bibitem{HNTW}
  S.~He, T.~Numasawa, T.~Takayanagi and K.~Watanabe,
  ``Notes on Entanglement Entropy in String Theory,''  JHEP {\bf 1505} (2015) 106  doi:10.1007/JHEP05(2015)106  [arXiv:1412.5606 [hep-th]].  


\bibitem{Cal}
  M.~M.~Caldarelli,
  ``Quantum scalar fields on anti-de Sitter space-time,''  Nucl.\ Phys.\ B {\bf 549} (1999) 499  [hep-th/9809144].  


\bibitem{BHS}
  S.~N.~Solodukhin,
  ``Entanglement entropy of black holes,''  Living Rev.\ Rel.\  {\bf 14} (2011) 8  [arXiv:1104.3712 [hep-th]].  

\bibitem{Za}
A.~B.~Zamolodchikov,
``Irreversibility of the Flux of the Renormalization Group in a 2D Field Theory,''  JETP Lett.\  {\bf 43} (1986) 730   [Pisma Zh.\ Eksp.\ Teor.\ Fiz.\  {\bf 43} (1986) 565].  

\bibitem{CaHu}
H.~Casini and M.~Huerta,
``A Finite entanglement entropy and the c-theorem,''  Phys.\ Lett.\ B {\bf 600} (2004) 142  [hep-th/0405111].  

\bibitem{Kabat}
  D.~N.~Kabat,
``Black hole entropy and entropy of entanglement,''  Nucl.\ Phys.\ B {\bf 453} (1995) 281  doi:10.1016/0550-3213(95)00443-V  [hep-th/9503016].  

\bibitem{CaHuMy}
H. Casini, M. Huerta, R. Myers, 
"Towards a derivation of holographic entanglement entropy,"   JHEP 1105 (2011) 036 [arXiv:1102.0440 [hep-th]]



\bibitem{DiDo}
D. E. Diaz, H. Dorn, 
"Partition functions and double-trace deformations in AdS/CFT," JHEP 0705 (2007) 046 [hep-th/0702163]

\bibitem{KlPuSa}
I. R. Klebanov, S. S. Pufu, B. R. Safdi, "F-Theorem without Supersymmetry," JHEP 1110 (2011) 038 [arXiv:1105.4598 [hep-th]]

\bibitem{GiKl}
S. Giombi, I. R. Klebanov, "Interpolating between a  and F," JHEP 1503 (2015) 117 [arXiv:1409.1937 [hep-th]]

\bibitem{GiKlPuSaTa}
S. Giombi, I. R. Klebanov, S. S. Pufu, B. R. Safdi, G. Tarnopolsky, "AdS Description of Induced Higher-Spin Gauge Theory," JHEP 1310 (2013) 016 [arXiv:1306.5242 [hep-th]]
 
\bibitem{ErHoOBWu}
J. Erdmenger, C. Hoyos, A. O'Bannon, J. Wu, "A Holographic Model of the Kondo Effect," JHEP 1312 (2013) 086  [arXiv:1310.3271 [hep-th]]

\bibitem{ErFlHoNeWu}
J. Erdmenger, M. Flory, C. Hoyos, M. N. Newrzella, J. M. S. Wu, "Entanglement Entropy in a Holographic Kondo Model," [arXiv:1511.03666 [hep-th]]




\end{thebibliography}
\end{document}